\documentclass[12pt]{article}

\usepackage{amsmath,latexsym,amssymb,epsfig,psfrag,axodraw,cite}

\newcommand{\be}{\begin{equation}}  
\newcommand{\ee}{\end{equation}}  
\newcommand{\bea}{\begin{eqnarray}}  
\newcommand{\eea}{\end{eqnarray}}

\begin{document}

\thispagestyle{empty}
\vspace*{.5cm}
\noindent
HD-THEP-05-06 \hfill 31 March 2005

\vspace*{1.9cm}

\begin{center}
{\Large\bf Radius Stabilization\\[0.3cm]
by Two-Loop Casimir Energy}
\\[2.3cm]
{\large G.~von Gersdorff$\,^a$ and A.~Hebecker$\,^b$}\\[.5cm]
{\it $^a$Department of Physics and Astronomy, Johns Hopkins University, 
3400 N Charles Street, Baltimore, MD 21218}\\[.2cm]
{\it $^b$Institut f\"ur Theoretische Physik, Universit\"at Heidelberg,
Philosophenweg 16 und 19, D-69120 Heidelberg, Germany}
\\[.4cm]
{\small\tt (\,gero@pha.jhu.edu\,,\,
hebecker@thphys.uni-heidelberg.de\,)}
\\[1.3cm]

{\bf Abstract}\end{center}
It is well known that the Casimir energy of bulk fields induces a non-trivial 
potential for the compactification radius of higher-dimensional field 
theories. On dimensional grounds, the 1-loop potential is $\sim 1/R^4$. 
Since the 5d gauge coupling constant $g^2$ has the dimension of length, the 
two-loop correction is $\sim g^2/R^5$. The interplay of these two terms
leads, under very general circumstances (including other interacting 
theories and more compact dimensions), to a stabilization at finite radius. 
Perturbative control or, equivalently, a parametrically large compact radius
is ensured if the 1-loop coefficient is small because of an approximate 
fermion-boson cancellation. This is similar to the perturbativity argument 
underlying the Banks-Zaks fixed point proposal. Our analysis includes 
a scalar toy model, 5d Yang-Mills theory with charged matter, the examination 
of $S^1$ and $S^1/Z_2$ geometries, as well as a brief discussion of the 
supersymmetric case with Scherk-Schwarz SUSY breaking. 2-Loop calculability 
in the $S^1/Z_2$ case relies on the log-enhancement of boundary kinetic 
terms at the 1-loop level.

\newpage
\section{Introduction}
Higher-dimensional field theories arise in the low energy limit of string- 
or M-theory, which is our best candidate for a theory of quantum 
gravity. Independently, compactified higher-dimensional models provide 
many interesting possibilities for the unification of known fields and 
interactions. Familiar examples are the appearance of 4-dimensional 
gauge theories as a manifestation of higher-dimensional diffeomorphism 
invariance or the unification of known bosons and fermions in 
higher-dimensional multiplets of supersymmetry (SUSY). Thus, we consider 
higher-dimensional compactified models a promising ingredient in 
possible physics beyond the standard model, which makes the further 
investigation of stabilization mechanisms for the compactification radius 
an interesting and potentially important subject.

One very generic ingredient in the dynamics of the compactification radius 
is the Casimir energy of massless bulk fields~\cite{Appelquist:1983vs}. As a 
simple example, consider 5d general relativity with a vanishing cosmological 
constant and a set of massless 5d fermions and bosons. Compactifying to 4d 
on an $S^1$ with physical volume $2\pi R$, one finds a flat 4d effective 
potential for the radion $R$. This flatness is lifted by the 1-loop Casimir 
energy which, on dimensional grounds, is $\sim 1/R^4$ and does not lead to 
a stable finite-$R$ solution.\footnote{Introducing 
a 
non-zero 5d cosmological constant $\Lambda_5$, a stable solution can be 
found by balancing the resulting $2\pi R\Lambda_5$ contribution against 
the Casimir energy. However, a positive 4d cosmological constant results
whose scale is set by $R$ and which is therefore generically too large. 
} 
Clearly, to overcome the $1/R^4$ behaviour which is too simple for 
stabilization, one has to introduce a mass scale into the potential, which 
can be achieved, for example, by considering warped 
compactifications~\cite{Garriga:2000jb}, massive bulk matter or brane
localized kinetic terms for bulk fields~\cite{Ponton:2001hq}.

In this paper, we point out that the required mass scale is, in fact,
generically present in the simplest interacting higher-dimensional
field theories, such as $\lambda\phi^4$ theories, gauge theories with
coupling constant $g$, or models with Yukawa interactions. In 5d the
above coupling constants are dimensionful, leading to a 2-loop Casimir
energy contribution $\sim\lambda/ R^5$ or $g^2/R^5$. Thus, radius
stabilization will generically arise at the 2-loop level by a
balancing of the $1/R^4$ and the $1/R^5$ contributions without the
need to invoke any extra effects or operators.\footnote{ Note that
this differs qualitatively from the results of the early discussion of
higher-loop Casimir stabilization in \cite{Antoniadis:1998sd} (see
below for more details).  } Generically, the compactification scale is
set by the lowest of the strong interaction scales of various 5d field
theories present in a given model.

We note that a different 2-loop stabilization mechanism was previously
considered in the context of 6d $\lambda\phi^3$ theory, where the
coupling is dimensionless and a logarithmic $R$-dependence arises at
the 2-loop level~\cite{Albrecht:2001cp}.  Furthermore, the possibility of 
2-loop stabilization based on the vanishing of the 1-loop contribution $1/R^4$ 
and a balancing of the $1/R^5$ and the $\ln(R)/R^5$ terms has been pointed 
out in \cite{DaRold:2003yi}.  Two-loop corrections to the 4d Casimir effect 
have been considered by many authors (see, e.g.,~\cite{Kay:1978zr}).

The paper is organized as follows. In Sect.~\ref{lf4}, the above idea is
illustrated using the simple example of 5d $\lambda\phi^4$ theory. We 
emphasize in particular that, by a judicious choice of the field content, 
stabilization at moderately large radii, $R\gg\lambda$, can be 
achieved, such that higher-loop corrections are negligible. This is 
similar to the way in which a perturbatively controlled non-trivial 
fixed point arises in the proposal of Banks and Zaks~\cite{Banks:1981nn}. 

Section~\ref{gt} extends the analysis to a 5d Yang-Mills theory with 
charged bosons and fermions. Amusingly, all 2-loop integrals reduce 
straightforwardly to the simple scalar case. Controlled 2-loop stabilization 
at large radius can be achieved, e.g., in the large $N$ limit of SU($N$) 
gauge theory with appropriate matter content. Furthermore, it is shown that 
our 2-loop stabilization mechanism extends straightforwardly to SUSY models 
with Scherk-Schwarz SUSY breaking. 

In Sect.~\ref{orb}, we provide a qualitative discussion of the 
phenomenologically more interesting cases of $S^1/Z_2$ and of the $S^1$ with 
3-branes. Since the finite and calculable 2-loop bulk contribution mixes 
with the 1-loop effect induced by brane operators with unknown coefficients, 
a complete predictivity just on the basis of the field content can not be 
achieved. In the generic case, the 2-loop effect considered here represents 
an ${\cal O}(1)$ correction to the previously discussed stabilization by 
brane-kinetic terms. We point out the interesting and natural limit of 
logarithmically enhanced brane-localized gauge-kinetic terms, which allows 
one to neglect the bulk-2-loop effect and to achieve full predictivity just
on the basis of the particle spectrum. 

A summary of our results as well as a discussion of possible further 
research directions, in particular the applicability to SUSY models on 
$S^1/Z_2$ and to the case of more than 5 dimensions, can be found in 
Sect.~\ref{conc}. 

The evaluation of the relevant loop integrals and non-trivial 
SUSY-based checks of the 2-loop gauge theory calculation
are described in two appendices.

\section{A $\lambda\phi^4$ example}\label{lf4}
Consider the 5d theory of Einstein gravity and massless real scalar with 
classical action 
\be
S=\int d^5x\sqrt{-g}\left(\frac{1}{2}\bar{M}_{P,5}^3{\cal R}_5+\frac{1}{2}
(\partial\phi)^2-\frac{\lambda}{4!}\phi^4\right)
\label{general}
\ee compactified on an $S^1$ with radius $R$. Clearly, a variation of
the metric background field $g_{55}$ is equivalently described by a
variation of the volume $2\pi R$. In the following, we will always use
a 5d Minkowski metric treating $R$ as our volume or radion degree of
freedom. It will not be necessary to perform a Weyl rescaling of the
metric to manifestly separate graviton and radion degrees of freedom in
the 4d effective theory.

Assuming that $\bar{M}_{P,5}\gg 1/\lambda$, we can consistently neglect 
gravitational interactions. On dimensional grounds, the effective potential 
for $R$ then reads
\be
V(R) = \frac{1}{R^4} \left( c^{(1)} + c^{(2)}\frac{\lambda}{R} + 
c^{(3)}\frac{\lambda^2}{R^2}+\dots\right).\label{CE}
\ee
where the $c^{(n)}$ are $n$-loop coefficients. Note that, even in the limit
of large 5d Planck mass, $c^{(1)}$ has to include the 5d graviton 
contribution. 

Radius stabilization can be achieved already at the 2-loop level. Indeed, 
if $c^{(1)}$ is negative and $c^{(2)}$ positive, the 2-loop potential is 
minimized at
\be
R =- \frac{5}{4}\frac{ c^{(2)}}{ c^{(1)}}\lambda\,.\label{apprmin}
\ee
However, the 4d cosmological constant at the minimum is negative. This can 
be remedied either by adding a 3-brane with appropriately tuned positive 
tension or a 5d bulk cosmological constant. In the second case, an extra 
$R$-dependent contribution to the 4d effective potential results, 
\be
V_{\rm cc}(R)=2\pi \Lambda_5 R.
\ee
Requiring both $V'(R)$ and $V( R)$ to vanish at the same point determines 
the precise value of $\Lambda_5$ and gives rise to a slightly shifted 
minimum at\footnote{
Strictly 
speaking, since we are not working in the Einstein frame, the equation of 
motion for $R$ is $V'+\pi\bar{M}_{P,5}^3{\cal R}_4=0$, where ${\cal R}_4$ 
is the 4d curvature. Demanding $V=0$ at the minimum also ensures that 
minimization of $V$ is equivalent to solving the equation of motion.}
\be
R=- \frac{6}{5}\frac{c^{(2)}}{ c^{(1)}}\lambda.
\ee

Unfortunately, assuming that $c^{(n)}={\cal O}(1)$, it is immediately
clear that higher-loop terms can not be neglected in the vicinity of
the above 2-loop minimum.\footnote{ 
We could, of course, improve our estimates by extracting appropriate
loop suppression factors from the coefficients $c^{(n)}$ using naive
dimensional analysis (see, e.g.,~\cite{Chacko:1999hg}). However, since
this would not affect our argument qualitatively, we do not enter such
a more detailed discussion. Alternatively, one can imagine that these
factors have already been absorbed in a redefined coupling.
}
This situation may, in fact, be generic. In this case we can not do more 
than express the justified hope that, in many models, higher-loop effects 
will significantly change but not destroy the minimum found at the 2-loop 
level. Our first conclusion is therefore that radius stabilization by 
higher-loop Casimir energy is presumably generic (in the sense of occuring in 
a large fraction of models) and the resulting compactification scale is 
of the order of the strong interaction scale of the most strongly coupled 
of the bulk field theories. 

However, in specific examples a quantitatively controlled minimum
based on the 2-loop approximation can occur. Indeed, in models where
$|c^{(1)}|\ll |c^{(2)}|$, the 2-loop minimum is at $R\gg \lambda$ and
higher-loop effects are suppressed (assuming that no undue enhancement
of the coefficients $c^{(n)}$ with $n\ge 3$ occurs). As a concrete
realization, consider the O$(N_s)$ symmetric generalization of the
above scalar $\lambda\phi^4$ lagrangian,
\be
\frac{\lambda}{4!}\phi^4\qquad\to\qquad\frac{\lambda}{4!}\left(\sum_{i=1}^{
N_s}\phi_i^2\right)^2\,,
\ee
together with $N_f$ fermions which are not (or only very weakly) coupled to 
the scalars. It is clear that, in this case, 
\be c^{(1)}\sim 4N_f-N_s-5\,, 
\ee
(which is the difference of fermionic and bosonic degrees of freedom,
including the 5 graviton polarizations) while
\be 
c^{(n)}\sim
N_s^n\qquad\mbox{for}\qquad n\ge 2\,.  
\ee
Thus, taking $N_s$ large while keeping $4N_f\!-\!N_s\!-\!5\sim {\cal
O}(1)$ and negative, one has $R\sim\lambda N_s^2$ at the 2-loop minimum
and all higher-loop terms are suppressed.

Finally, we now want to fill in the explicit numbers for the first two
loop coefficients used above. As already mentioned, $c^{(1)}$ is 
proportional to the difference of on-shell fermionic and bosonic degrees of 
freedom of the 5d theory,
\be
c^{(1)}=\left(N_{\rm fermions}-N_{\rm bosons}\right)c_0^{(1)}\,,\label{c1}
\ee
where~\cite{Appelquist:1983vs,Ponton:2001hq} (see also Appendix~A)
\be
c^{(1)}_0\equiv\frac{3\,\zeta(5)}{(2\pi)^6}\,.\label{c10}
\ee
The 2-loop coefficient for a single scalar is due to the ``figure-8'' 
diagram, which has previously been derived using the winding mode 
expansion~\cite{DaRold:2003yi}. In our context, it is crucial that the 
tree-level masslessness in 5d is maintained by an appropriate 1-loop 
counterterm.\footnote{There 
is 
however a finite (nonlocal) positive correction to the scalar mass 
squared of the 4d zero mode.
} 
The result reads (for explicit calculations see Appendix~A)
\be
c^{(2)}=\frac{\zeta(3)^2}{8(2\pi)^9}\,.\label{c2} 
\ee
Going from a single 
scalar to the O$(N_s)$-symmetric model, the coefficient $c^{(2)}$ of 
Eq.~(\ref{c2}) has to be multiplied by $(N_s^2+2N_s)/3$. For our simple
scalar example to work, it is important that $c^{(2)}>0$.

Thus, a quantitatively controlled 2-loop minimum arises from the potential
\be
V(R)=\frac{1}{(2\pi)^6R^4}\left\{3\,\zeta(5)(4N_f-N_s-5)+\frac{\zeta
(3)^2}{24(2\pi)^3}(N_s^2+2N_s)\,\frac{\lambda}{R}\right\}\,,\label{vn}
\ee
in the specific large-$N_s$ limit described above. 

While this simple analysis shows that it is quite easy to achieve
stabilization in a scalar $\phi^4$ theory, a more interesting and realistic 
case is that of a 5d gauge theory. In particular, the masslessness 
assumption introduced above, which is unnatural for an interacting scalar, 
will be natural for gauge bosons and charged fermions.

\section{Gauge theory}\label{gt}

We now turn to the case of a 5d gauge theory compactified on $S^1$ with the 
action
\be
S=\int d^5x\sqrt{-g}\left(\frac{1}{2}\bar{M}_{P,5}^3{\cal R}_5-\frac{1}{2g^2}
\mbox{tr}_f\left(F_{MN}F^{MN}\right)+(D_M\Phi)^\dagger(D^M\Phi)+
\bar{\Psi}iD\!\!\!\!/\,\,\Psi\right)\,.
\ee
Here the bosonic and fermionic matter fields transform in some representation 
$r$ of the gauge group $G$, the field strength is $F_{MN}=-i[D_M,D_N]$
with $D_M=\partial_M+iA_M$, and the trace is in the fundamental 
representation with generators normalized by 2$\,$tr$_f({\bf T}^a{\bf T}^b)
=\delta^{ab}$. 

The 2-loop vacuum energy contributions come from the diagrams
depicted in Figs.~\ref{gauge} and \ref{matter}. It turns out that, by simple 
algebraic manipulations, they can all be reduced to the 2-loop 
integral encountered in the scalar model above. Therefore, the complete 
result can be expressed in terms of the constant 
\be
c^{(2)}_0=\frac{\zeta(3)^2}{(2\pi)^9}\,\dim(G) 
\ee
where $\dim(G)$ is the dimension of the gauge group (the $1/8$ of 
Eq.~(\ref{c2}) is a symmetry factor characteristic of the $\lambda\phi^4$ 
model). 

\begin{figure}[tb]
\begin{center}
\SetScale{.75}
\begin{picture}(50,100)(25,0)
\PhotonArc(50,75)(23,14,374){3}{13}
\PhotonArc(50,25)(23,194,554){3}{13}
\end{picture}
\hspace{2cm}
\begin{picture}(50,100)(25,0)
\PhotonArc(50,50)(25,14,374){3}{13}
\Photon(25,50)(75,50){3}{5}
\end{picture}
\hspace{2cm}
\begin{picture}(50,100)(25,0)
\DashCArc(50,50)(25,-85,270)2
\Photon(25,50)(75,50){3}{5}
\end{picture}
\end{center}
\caption{\em Two-loop diagrams from the gauge sector.}
\label{gauge}
\end{figure}
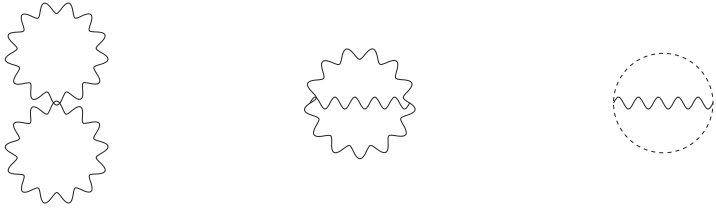

\begin{figure}[tb]
\begin{center}
\SetScale{.75}
\begin{picture}(50,100)(25,0)
\CArc(50,50)(25,14,374)
\Photon(25,50)(75,50){3}{5}
\end{picture}
\hspace{2cm}
\begin{picture}(50,100)(25,0)
\PhotonArc(50,75)(24,14,374){3}{13}
\DashCArc(50,25)(24,-85,270)4
\end{picture}
\hspace{2cm}
\begin{picture}(50,100)(25,0)
\DashCArc(50,50)(25,-85,270)4
\Photon(25,50)(75,50){3}{5}
\end{picture}
\end{center}
\caption{\em Two-loop diagrams from the matter sector.}
\label{matter}
\end{figure}
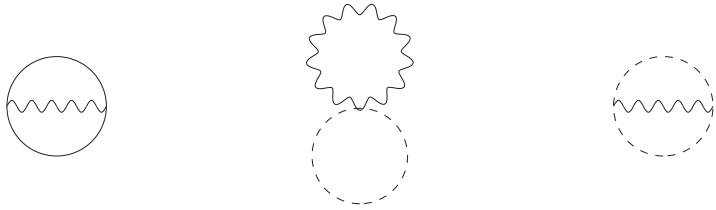

Including the symmetry factors and numerator algebra of the various diagrams 
and accounting for the trace normalization through the constant $T(r)$, 
defined by tr$_r({\bf T}^a{\bf T}^b)=T(r)\delta^{ab}$ (which is also known 
as the Dynkin index of the representation $r$), one finds
\be
\begin{array}{rclcl}
c^{(2)}_{\rm vector}/c^{(2)}_0&=& \frac{1}{4}d(d-1)\,T(a) 
-\frac{3}{4}(d-1)\,T(a) + \frac{1}{4} \,T(a) 
&=& +\frac{9}{4}\,T(a)\,,\\ \\
c^{(2)}_{\rm scalar}/c^{(2)}_0&=& d\,T(r)-\frac{3}{2}
T(r)&=&+\frac{7}{2}\,T(r)\,,\\ \\
c^{(2)}_{\rm fermion}/c^{(2)}_0&=&(2-d)T(r)&=& -3\,T(r)\,.
\end{array}
\label{2loopgauge}
\ee
The sum of these coefficients defines $c^{(2)}$ of Eq.~(\ref{CE}) (with 
$\lambda$ replaced by $g^2$) and
thereby the 2-loop contribution to the Casimir energy arising from gauge 
fields and gauged matter. To facilitate comparison with other calculations, 
we have made explicit the $d$-dependence before setting $d=5$ and specified 
the contributions arising from the separate Feynman gauge diagrams. 
Specifically, the three contributions to $c^{(2)}_{\rm vector}$ come from the 
``figure 8'', the ``setting sun'' diagram, and the ``setting sun'' diagram 
with ghosts (in this order, cf. Fig.~\ref{gauge}). Analogously, the two 
contributions to $c^{(2)}_{\rm scalar}$ arise from the ``figure 8'' and the 
``setting sun'' diagram (cf.~Fig.~\ref{matter}). A non-trivial check of these
results based on the vanishing of the Casimir energy in models with unbroken 
SUSY can be found in Appendix~B. 

One can see that in a pure gauge theory, one again encounters precisely the 
situation outlined above: $c^{(1)}$ is negative while $c^{(2)}$ is positive
so that 2-loop stabilization is automatic. In order to maintain 
perturbativity, we need to reduce the one-loop coefficient without affecting 
the other $c^{(n)}$. This is most easily achieved by considering large 
groups (e.g. SU($N$) with $N$ large, where $c^{(n)}\sim N^{1+n}$) and adding
fermions uncharged under the gauge group to reduce $c^{(1)}$ to ${\cal O}
(1)$. This will lead to $R\sim g^2N^3$ at the 2-loop minimum and result in 
a relative suppression of the $n$-loop ($n>2$) contribution near this minimum 
by $1/N^{2n-4}$. In fact, this situation may arise fairly naturally since 
higher-loop coefficients are dominated by the most strongly coupled gauge 
group factor, so that it is sufficient to require the relevant fermions to be 
neutral only under this part of the group. We have thus seen that it 
is easy to stabilize the $S^1$ radius $R$ in a controlled fashion. Of 
course, to build a realistic model one would have to include branes or to 
consider more sophisticated geometries allowing for chiral fermions. 

We close this section by commenting on a possible SUSY version of this 
scenario. Consider a model containing a 5d supergravity multiplet, $N_V= 
\dim(G)$ vector multiplets defining a super Yang-Mills theory with gauge group 
$G$, and $N_H$ hypermultiplets in a representation $r$ of $G$. Let us break
SUSY from ${\cal N}=2$ to ${\cal N}=0$ by introducing Scherk-Schwarz boundary 
conditions on the $S^1$~\cite{Scherk:1978ta}. The effect of this breaking is a 
mass shift for gauginos, gravitinos and hyperscalars. Their Kaluza-Klein (KK) 
masses become $m_n R=n+\omega$, where the real parameter $\omega$ is known as 
the Scherk-Schwarz parameter. Already at this point it is clear that our
stabilization mechanism is qualitatively unchanged: The Scherk-Schwarz 
parameter is dimensionless and our basic formula, Eq.~(\ref{CE}), remains 
valid. Of course, the coefficients $c^{(i)}$ are now functions of $\omega$
(which vanish for $\omega=0$). 

The 1-loop contribution to the Casimir energy is specified 
by~\cite{Antoniadis:1998sd,Delgado:1998qr,vonGersdorff:2003rq}
\be
c^{(1)}_{\rm SS}=\frac{12}{(2\pi)^6}(N_H-N_V-2)\left\{
\zeta(5)-\zeta_\omega(5)
\right\}\,,
\label{susy1loop}
\ee
where we have defined the function $\zeta_\omega(n)$ by
\be
\zeta_\omega(n)=\sum_{k=1}^{\infty} \frac{\cos(2 \pi k \omega)}{k^n}\,.
\ee
Since $\zeta_\omega(n)\leq \zeta_0(n)=\zeta(n)$, this contribution is 
negative as long as $N_V+2>N_H$. We can now balance matter and gauge 
multiplets in exactly the same way as in the non-SUSY case to ensure 
$c^{(1)}={\cal O}(1)$. For the 2-loop contribution of the vectormultiplet, 
we use the results of Appendix~B but evaluate the corresponding gaugino 
integrals with the shifted masses to obtain
\be
c^{(2)}_{\rm SS\ vector}=\frac{4}{(2\pi)^9}
N_V T(a)
\left\{\zeta(3)-\zeta_\omega(3)\right\}^2.
\ee
Likewise, the hypermultiplet contribution reads\footnote{
The relation $c^{(i)}_{\rm SS\ vector}+c^{(i)}_{\rm SS\ hyper}=0$ for an 
adjoint hypermultiplet representation reflects the fact these ${\cal N}=2$ 
multiplets combine into ${\cal N}=4$ vector multiplets. The Scherk-Schwarz 
mechanism leaves ${\cal N}=2$ SUSY unbroken and the Casimir energy remains 
zero.}
\be
c^{(2)}
_{\rm SS\ hyper}=-\frac{4}{(2\pi)^9}
N_V T(r)
\left\{\zeta(3)-\zeta_\omega(3)\right\}^2.
\ee
Since, as in the non-SUSY case, the vector multiplet contribution is 
positive, our 2-loop stabilization mechanism remains effective in this
simple SUSY model as long as there is not too much matter charged under the 
most strongly coupled gauge group factor.

\section{Compactifications with branes and fixed-points}\label{orb}
In this section, we focus on the 5d gauge theory case discussed in 
Sect.~\ref{gt} since it is presumably more likely to be part of 
phenomenologically interesting theories than the scalar toy model of 
Sect.~\ref{lf4}. However, most of our discussion applies, qualitatively, 
also to the scalar case and presumably to many other 5d models with 
dimensionful couplings and corresponding 2-loop Casimir stabilization. 

As already mentioned above, the pure $S^1$ case discussed up to now can not 
give rise to realistic models since the 4d particle spectrum is necessarily
vector-like. A simple way of embedding this type of $S^1$ stabilization in 
a realistic construction is to add a 3-brane on which matter fields can 
live.\footnote{
As
already mentioned earlier, a brane localized cosmological constant can be 
used to tune the effective 4d cosmological constant to zero. Its interplay 
with a possible bulk cosmological constant can also affect stabilization (for 
more options along these lines see, e.g.,~\cite{Hofmann:2000cj}). Here, we 
require the bulk cosmological constant to be small because of some symmetry
principle (e.g. bulk SUSY) and treat the sum of brane cosmological constants 
as an unknown parameter to be fixed by the requirement of 4d flatness. 
} 
To apply our stabilization mechanism without any modification, one 
could simply assume that the brane fields are not charged under the bulk 
gauge theory responsible for the stabilizing 2-loop Casimir energy. 

If one does not make this assumption, the charged brane fields
generate, at the 1-loop level, brane localized gauge-kinetic terms for
the bulk gauge theory. Such brane-kinetic terms contribute to the
1-loop Casimir energy induced by bulk fields. We can view this as a
2-loop effect including both brane and bulk fields. It is easy to
convince oneself that this contribution is not parametrically
suppressed relative to the bulk 2-loop effect calculated in
Sect.~\ref{gt}. Thus, we are precisely in the situation of 1-loop
Casimir stabilization with brane kinetic terms considered
in~\cite{Ponton:2001hq}. Our 2-loop calculation is then simply a
finite ${\cal O}(1)$ correction to this stabilization
mechanism. However, since the coefficient of the relevant
brane-kinetic term is UV-sensitive, its value after renormalization is
in essence just a new parameter of the model. The precise result of
the 2-loop calculation is then only meaningful if one has some
first-principles knowledge about the coefficient of the brane kinetic
term. This will, in general, require a UV completion as it is
provided, e.g., by a string orbifold model.

A very similar situation arises in an $S^1/Z_2$ geometry. Here, not
only brane-localized charged fields but also the bulk gauge and matter
fields themselves induce brane-localized gauge-kinetic terms. This is,
in fact, a familiar and well-studied effect in the context of
higher-dimensional grand unified theories (GUTs) with symmetry
breaking by brane-localized Higgs fields and in orbifold
GUTs~\cite{Nomura:2001mf,Hebecker:2001wq}. It corresponds to the
statement that a (modified) logarithmic running of the gauge couplings
continues above the compactification scale, which can be understood as
the running of the coefficient of a brane localized $F^2$
term~\cite{Hebecker:2001wq}. 

In fact, it is easy to check that, even in the simple scalar model, the 
2-loop vacuum energy in an $S^1/Z_2$ compactification has (in contrast to 
the pure $S^1$ case) a logarithmic divergence that can be absorbed into a 
brane-localized $\phi\,\partial_5^2\, \phi$ operator. This is analogous to 
the boundary term $F_{\mu\nu}F^{\mu\nu}$ (with $\mu,\nu\in\{0,1,2,3\}$)
induced in gauge theories with even boundary conditions for $A_\mu$. However, 
while in the gauge theory case only the 4d-part of the gauge-kinetic term is 
corrected ($F_{5\mu}$ being zero at the boundary), in the scalar case the 
correction only affects the $\partial_5$ part (the 1-loop self energy diagram 
being momentum-independent). 

The apparent loss of predictivity associated with the UV sensitive
coefficients of brane-localized kinetic terms is not as severe as one
might naively think. The reason is the logarithmic enhancement of such
terms associated with their logarithmic divergence. Indeed, the bulk
gauge theory has a strong interaction scale associated with the 5d
gauge coupling, $M\sim 24\pi^3/g^2$~\cite{Chacko:1999hg}. It is
natural to assume that, at this scale, the brane-localized $F^2$ term
has an ${\cal O}(1)$ coefficient.  Running down to the
compactification scale $M_c=1/R$, one obtains a log-enhanced
coefficient $\sim\ln(M/M_c)=\ln(24\pi^3R/g^2)$. This is dominant with
respect to the unknown ${\cal O}(1)$ initial value. The corresponding
log-enhanced 1-loop Casimir effect contribution of the brane operator
is also dominant with respect to the true 2-loop effect calculated in
Sect.~\ref{gt}. Thus, the leading piece of the $g^2/R^5$ contribution
is calculable on the basis of the low-energy field content of the
model. The parametric behaviour is, in fact, not a pure power of $R$
but includes the logarithmic $R$ dependence of the coefficient. The 4d
vacuum energy of a 5d gauge theory compactified on $S^1/Z^2$ thus has
the form~\footnote{ As already mentioned in the Introduction,
higher-loop radius stabilization was also discussed in
\cite{Antoniadis:1998sd}, but in a very different approach. We
understand that Ref.~\cite{Antoniadis:1998sd} treats the 4d coupling
as fundamental and $R$-independent, resulting in loop corrections
$\sim\ln(R)/R^4$.  By contrast, we consider the 5d coupling as fundamental,
which implies a 4d coupling $\sim 1/R$ and hence a vacuum energy loop
correction $\sim\ln(R)/R^5$.  

Furthermore, a potential similar to Eq.~(\ref{v4b}) was derived 
in~\cite{DaRold:2003yi} for a scalar model with Yukawa couplings to 
brane fields. Assuming an exact cancellation of the 1-loop contributions 
from scalars and fermions, stabilization was achieved by a balancing of 
the $1/R^{5}$ and $\ln(R)/R^5$ terms of the 2-loop result.
}
\be
V_4=\frac{1}{R^4}\left(c^{(1)}+c^{(2)}\frac{g^2\ln(MR)}{R}\right)\,,
\label{v4b}
\ee
and can, as explained earlier, give rise to quantitatively controlled
radius stabilization at $R\sim g^2N^3$ in the case of an SU($N$) gauge
theory.

To be more specific, focus on 5d SU($N$) gauge theory on $S^1/Z_2$ (where the 
gauge group is not broken by the orbifolding) with $N_f$ uncharged bulk 
fermions and with charged brane fermions in a representation $r$ at the 
$x^5=0$ boundary. The logarithmic running above the compactification scale 
induces an effective brane-localized $F^2$ term 
\be
{\cal L}\supset -\frac{1}{96\pi^2}\mbox{tr}(F_{\mu\nu}F^{\mu\nu})
\ln(MR)\left[\delta(x^5)\left(8T(r)-\mbox{$
\frac{23}{4}$}T(a)\right)+\delta(x^5-\pi R)\left(-\mbox{$\frac{23}{4}$}T(a)
\right)\right]\label{bt}
\ee
at scale $M_c$. The calculations leading to this formula are explained in 
detail in Sect.~3 of~\cite{Hebecker:2002vm} and can also be extracted from 
the earlier references~\cite{Nomura:2001mf,Hebecker:2001wq}. Here the strong 
interaction scale is $M\simeq 24\pi^3/(g^2N)$, accounting for the large-$N$
enhancement of higher loops in SU($N$) gauge theory. As far as the 
1-loop Casimir energy calculation using a summation of KK modes 
is concerned, the effect of this contribution can be summarized by an extra 
contribution to the momentum-space effective action, 
\be
\frac{\pi R}{2g^2}\left(k^2\delta_{\mu\nu}-k_\mu k_\nu\right)\quad\to
\quad\left(\frac{\pi R}{2g^2}+2b\right)\left(k^2\delta_{\mu\nu}-k_\mu k_\nu
\right)\,,
\ee
to be used for even modes (i.e. the $A_\mu$ modes) only. Here $b$ is defined 
by minus the sum of the two coefficients of the brane-localized $F^2$ terms 
as given in Eq.~(\ref{bt}) and $k$ is the euclidean 4-momentum. The extra 
factor of 2 arises since the even higher KK modes are twice more sensitive 
to brane operators than the zero mode. 

Based on this, the 1-loop contribution to the 4d potential induced by the 
brane-kinetic terms reads
\bea
V^{\rm brane}(R)&=&\frac{3\dim(G)}{2}\sum_{n=1}^{+\infty}\int\frac{d^{d-1}k}
{(2\pi)^{d-1}}\left\{\ln[(1+b')k^2+(nM_c)^2]-\ln[k^2+(nM_c)]\right\}
\nonumber \\ \nonumber \\
&\simeq &\frac{3\dim(G)}{2}\sum_{n=1}^{+\infty}\int\frac{d^{d-1}k}
{(2\pi)^{d-1}}\,\frac{b'\,k^2}{k^2+(nM_c)^2}\,.\label{vbr}
\eea
Here $b'=2b/(\pi R/2g^2)$ and the summation is only over even non-zero 
(cosine) modes of the KK mode expansion on the $S^1$ compact space. The 
prefactor of 3 accounts for the 3 on-shell degrees of freedom of a massive 
vector. Note that we have simply interpreted the boundary operator as a 
correction to the energy of each separate leading-order KK mode. (This 
shifted KK mode spectrum has been derived in a different way in Appendix B 
of~\cite{Cheng:2002iz}.) To be more precise, one would have to 
re-diagonalize the quadratic-order Hamiltonian after inclusion of the 
boundary terms and then work with the modified KK mode 
expansion~\cite{Ponton:2001hq}. However, as long as $\ln(MR)\ll MR$, we are 
entitled to linearize in the boundary term and our simplified treatment 
is sufficient. 

Using the calculational techniques of Appendix~A, the above formula is
evaluated to give the brane-induced correction (equivalent to the
dominant part of the 2-loop Casimir energy)
\be V^{\rm brane}(R)=\frac{36\zeta(5)g^2\dim(G)\,b}{(2\pi)^6 R^5\pi}\,, 
\ee
leading to the final result\footnote{The factor of $\frac{1}{2}$ in
the one-loop contribution is due to the orbifolding, since half of the
modes are projected away (cf.~also the summation in Eq.~(\ref{vbr})).}
\be
V(R)=\frac{3\zeta(5)}{(2\pi)^6R^4}\left\{\frac{4N_f-3(N^2\!-\!1)-5}{2}
+\frac{N^2\!-\!1}{(2\pi)^3}\,\left[\,8T(r)-\frac{23}{2}T(a)\,\right]\,
\frac{g^2\ln(MR)}{R}\right\}\,.  
\ee
It is now easy to arrange for the coefficient of the $1/R^4$ piece to
be negative and ${\cal O}(1)$ while keeping the $1/R^5$ term positive
and ${\cal O}(g^2N^3)$ such that, as before, controlled 2-loop
stabilization at a moderately large radius is achieved.\footnote{To
turn over the sign in the second term, we need to introduce a certain
amount of brane fermions. Note also that assigning negative parities
to some gauge fields -- in other words, breaking the gauge
group by orbifolding -- would reduce the contribution of the gauge fields
to the brane kinetic terms and can even flip its sign.} Compared with
the $S^1$ case, where $MR\sim N^2$ we now have $MR/\ln(MR)\sim
N^2$. Using asymptotic properties of the Lambert $W$ function (see,
e.g.,~\cite{lambert}) the solution for $N$ large can be written as
\be
MR\sim N^2 \nu\,\bigl\{1+{\cal O}(\ln \nu/\nu)\bigr\},\qquad \nu=\ln N^2\,,
\ee
thus giving an additional enhancement factor $\ln N^2$.

\section{Conclusions}\label{conc}

We have presented a mechanism stabilizing the size $R$ of an extra
dimension compactified on $S^1$ or on the orbifold $S^1/\mathbb Z_2$ which 
is based on the presence of dimensionful couplings -- a generic feature of 
field theoretic models with $d>4$. We have shown that, by balancing 
the 1- and 2-loop contributions to the Casimir energy, a perturbatively 
controlled minimum at moderately large values of $R$ can be realized.

In the case of scalar massless $\phi^4$ theory on $S^1$, the Casimir
energy is calculable and finite as long as the masslessness is
enforced as a renormalization condition. The 1-loop contribution scales like 
$\sim -R^{-4}$, while the 2-loop effect of the scalar self-interaction 
gives $\sim+\lambda R^{-5}$, thus producing a nontrivial minimum. In order 
to ensure that the result is perturbatively controlled, we have added weakly 
coupled fermions. Their effect is to reduce the numerical factor of the 
1-loop term while only very mildly affecting the higher-loop contributions. 
This shifts the minimum to larger values of $R$.

The situation is basically the same in the case of a gauge theory. As before, 
the purely bosonic theory already produces a 2-loop minimum in the radion 
potential, while the inclusion of fermions uncharged under the most strongly 
coupled gauge group factor ensures the perturbativity of the result. We have 
also shown that in a SUSY extension of this scenario (with SUSY-breaking \`a 
la Scherk-Schwarz), the mechanism works without modification. 

The above calculable two-loop stabilization is modified but not destroyed 
if, instead of the circle, the orbifold $S^1/\mathbb Z_2$ is considered. 
The crucial new feature of this situation are brane-kinetic terms for the 
gauge fields, which are generated in the 1-loop effective action. Without the 
knowledge of the underlying UV physics one cannot predict the corresponding 
counterterm at the cutoff scale $M$. However, in calculating the potential 
of $R$ for large values of the radius, $R\gg M^{-1}$, one effectively 
integrates out the physics down to that scale, and the logarithmic running of 
the brane-kinetic term dominates over its unknown initial value. The dominant 
2-loop effect can then be obtained by evaluating the 1-loop integral in
the presence of this log-enhanced brane-kinetic term. The new contribution
scales like $\sim g^2\ln (R\,M)/R^5$ and is calculable on the basis of the 
bulk and brane field content of the model. We emphasize that this 
log-divergent part of the 2-loop calculation dominates over the remaining
(finite) 2-loop effects (scaling as $\sim 1/R^5$) as well as the one-loop 
effect of the unknown brane-kinetic term (whose leading effect for large $R$ 
is also $\sim 1/R^{5}$).

Although we have focussed exclusively on 5d models, we note that our 
mechanism is equally suitable for other higher-dimensional manifolds with 
a single unstabilized modulus. This is because such theories look very 
similar from a 4D point of view, the main difference being a modified 
KK-mass spectrum which depends on the value of the modulus. Various 
possibilities for stabilizing such a single modulus have recently been 
discussed in the context of the KKLT proposal~\cite{Kachru:2003aw}, and we 
believe that Casimir energies will be relevant in this context. This has, 
in fact, very recently been discussed in the context of the 1-loop effect 
of massive vector fields in~\cite{Dudas:2005vn}, and one can expect that the 
higher-loop contributions discussed here will also play an important role in 
the further study of moduli stabilization in more complete, higher-dimensional
scenarios. 

As a simple and specific example, we briefly 
consider $S^n$ compactifications with the volume undetermined by the 
Einstein equations. The Casimir energy coming from a self-interacting scalar
field will take the same general form as in Eq.~(\ref{CE}), with
the powers of $R$ modified according to the dimensionality of the coupling and 
the $c^{(i)}$ to be calculated from the appropriate KK spectrum.
The coefficient $c^{(1)}$ has been calculated for the case of a sphere
(without additional infinite dimensions) in Ref.~\cite{Elizalde:1993ud} 
for $n=1,2,3,4$ and found to be negative. Moreover, it is clear that 
$c^{(2)}$ is always positive since the integral appears squared. Thus, 
if the inequality $c^{(1)}<0$ survives the transition from $S^n$ to 
$M_4\times S^n$ for $n>1$ (as it does for $n=1$), our stabilization 
mechanism extends straightforwardly to these geometries. In any case, we
can be confident that many examples of 2-loop (or higher-loop) 
Casimir stabilization exist within the rich class of models with $n>1$ 
compact dimensions where the total volume is a modulus at tree level. 

Our final point concerns SUSY theories on $S^1/\mathbb
Z_2$. The prototype SUSY breaking mechanism on $S^1$, the Scherk-Schwarz
mechanism\cite{Scherk:1978ta}, extends to this case. It can equivalently be
described by giving a vacuum expectation value to the $F$ component of
the radion superfield (the chiral superfield whose lowest component contains 
$R$)~\cite{Marti:2001iw,vonGersdorff:2001ak}. Indeed, the resulting
tree level action corresponds to no-scale supergravity and the
potential for $R$ is completely flat.  Stabilization at 1-loop has
previously been studied by use of bulk mass terms for
hypermultiplets~\cite{Ponton:2001hq,Luty:2002hj,vonGersdorff:2003rq}, brane 
kinetic terms~\cite{Ponton:2001hq,Rattazzi:2003rj} or massive vector
multiplets~\cite{Dudas:2005vn}.  The mechanism we described in
Sec.~\ref{orb} seems to be quite suitable for SUSY $S^1/\mathbb Z_2$ models 
without 5d masses. The 1-loop contribution is given by 
$\frac{1}{2}$ of the $S^1$ result obtained in this paper.  We expect the 
dominant 2-loop effect to be the log-enhanced brane-kinetic terms, the 
same as in the non-SUSY case. However, a detailed analysis of this 
scenario, of its interplay with the other stabilization mechanisms mentioned 
above, or even the construction of a realistic SUSY model go beyond the scope 
of the present paper and are left to future research.

\noindent
{\bf Acknowledgements}: We would like to thank Roberto Contino, Alex
Pomarol, Mariano Quir\'os, Riccardo Rattazzi and Enrico Trincherini
for useful comments and discussions. GG is supported by the Leon
Madansky Fellowship and by NSF Grant P420-D36-2051.

\section*{Appendix A: Basic 1- and 2-loop integrals}
\setcounter{equation}{0}\renewcommand{\theequation}{A.\arabic{equation}}
We begin by rederiving the known 1-loop vacuum energy for a real scalar 
field on $R^4\times S^1$ in dimensional regularization. The 1-loop effective 
potential of the 5d euclidean theory is given by the $d\to 5$ limit of
\be
V_5^{(1)}(R)=\frac{1}{2R}\sum_{n=-\infty}^{+\infty}\int\frac{d^{d-1}k}
{(2\pi)^d}\ln[k^2+(nM_c)^2]\,,
\ee
where $M_c=1/R$. Using the fact that the $d$-dimensional integrals of any 
power of $k^2$ and of $\ln(k^2)$ vanish in dimensional regularization, we 
have 
\bea
V_5^{(1)}(R)&=&\,\,\frac{M_c}{2}\int_0^{M_c^2}dM^2\int\frac{d^{d-1}k}{(2\pi)^d}
\sum_{n=-\infty}^{+\infty}\frac{n^2}{k^2+(nM)^2}
\\
&=&-\frac{M_c}{2}\int_0^{M_c^2}dM^2\int\frac{d^{d-1}k}{(2\pi)^d}\,
\frac{k^2}{M^4}\,\,\frac{\pi\mbox{coth}(\pi|k|/M)}{|k|/M}\,.
\eea
Appealing again to the fact that the $d$-dimensional integral of a pure power 
vanishes, $\mbox{coth}(\pi|k|/M)$ can be replaced by $\mbox{coth}(\pi|k|/M) 
\!-\!1$, making the $d^{d-1}k$ integral finite and allowing one to take the 
limit $d\to 5$. The resulting 4d effective potential is
\be
V_4^{(1)}(R)=2\pi R\,V_5^{(1)}(R)=-\frac{3\zeta(5)}{(2\pi)^6R^4}\qquad 
\mbox{with}\qquad\zeta(5)=1.0369...\,,
\ee
in agreement with Eqs.~(\ref{c1}) and (\ref{c10}).

Higher-loop corrections to the effective action (in particular to 
$-V_5^{(1)}$) are given by the sum of all one-particle-irreducible vacuum 
diagrams of the relevant field theory. As already pointed out in the main 
text, the 2-loop correction is essentially just the ``figure 8'' diagram. 
To be more precise, one has to add to the lagrangian the mass counterterm 
of the uncompactified 5d theory ensuring that the scalar remains massless 
at the 1-loop level. This counterterm becomes part of a tadpole-like 1-loop 
vacuum diagram, which is part of the 2-loop correction. However, in 
dimensional regularization the mass counterterm vanishes and the 2-loop 
correction (for a single scalar field) simply reads 
\be
V_5^{(2)}=\frac{\lambda}{8}\,I^2\,,\qquad\mbox{where}\qquad
I=\frac{1}{R}\sum_{n=-\infty}^{+\infty}\int\frac{d^{d-1}k}
{(2\pi)^d}\,\,\frac{1}{k^2+(nM_c)^2}\,.
\ee
Here the prefactor contains a symmetry factor $(1/8)$ and the tadpole
integral $I$ is evaluated as the $d\to 5$ limit of
\be
I=\int\frac{d^{d-1}k}{(2\pi)^d}\,\,\frac{\pi\mbox{coth}(\pi|k|/M)}{|k|}=
\frac{\zeta(3)}{(2\pi)^5R^3}\qquad\mbox{with}\qquad \zeta(3)=1.2021...
\ee
This completes the derivation of Eq.~(\ref{c2}) and thereby of 
Eq.~(\ref{vn}).

\section*{Appendix B: SUSY checks of gauge theory results}
\setcounter{equation}{0}\renewcommand{\theequation}{B.\arabic{equation}}
To verify the results of our 2-loop gauge theory calculation, we have 
performed checks based on the vanishing of the Casimir energy in theories 
with unbroken SUSY in 5 and 4 dimensions. 

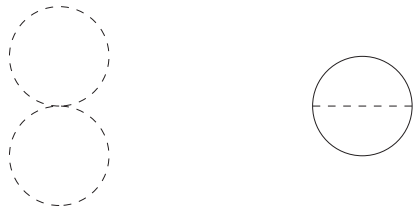
\begin{figure}[ht]
\begin{center}
\SetScale{.75}
\begin{picture}(50,100)(25,0)
\DashCArc(50,75)(25,-85,270)4
\DashCArc(50,25)(25,-85,270)4
\end{picture}
\hspace{2cm}
\begin{picture}(50,100)(25,0)
\CArc(50,50)(25,-90,270)
\DashLine(25,50)(75,50)4
\end{picture}
\end{center}
\caption{\em Two-loop diagrams from Yukawa couplings and scalar 
self-interactions.}
\label{yukawa}
\end{figure}

Specifically, we can consider a 5d super Yang-Mills (SYM) theory which, in 
addition to the gauge fields, includes a real adjoint scalar $\Sigma$ and a 
Dirac gaugino. To calculate the 2-loop effect in this theory, coupled to a 
hypermultiplet (a Dirac fermion and two complex scalars) in a representation 
$r$ of the gauge group, we need the contributions coming from the additional 
Yukawa- and scalar self-interaction diagrams shown in Fig.~{\ref{yukawa}}.
For the explicit lagrangian see, e.g.,~\cite{Fayet:1985ua}. 

The three ``setting-sun'' diagrams coming from the Yukawa couplings
yield
\begin{eqnarray}
c^{(2)}_{\rm gaugino-fermion-scalar} &=&-8\, c^{(2)}_0\, T(r)\,, \nonumber\\
c^{(2)}_{\rm fermion-\Sigma}&=&- c^{(2)}_0\, T(r)\,,  \label{ss}\\
c^{(2)}_{\rm gaugino-\Sigma}&=&- c^{(2)}_0\, T(a)\,.\nonumber
\end{eqnarray}
Furthermore, in the scalar sector there are two new ``figure-8'' diagrams 
coming from the coupling of $\Sigma$ to the scalar and of the scalar to 
itself, giving  two new contributions
\begin{eqnarray}
c^{(2)}_{{\rm scalar}-\Sigma}&=&2\,c^{(2)}_0\, T(r)\,,\nonumber\\
c^{(2)}_{\rm scalar-scalar}&=&3 \,c^{(2)}_0\, T(r)\,.\label{f8}
\end{eqnarray}
Finally, the gaugino and $\Sigma$ as well as the hypermultiplet fermion 
and scalar are charged under the gauge groups. The corresponding 2-loop 
corrections can be directly read off from Eq.~(\ref{2loopgauge}). The two
bosonic contributions are
\begin{eqnarray}
c^{(2)}_{\rm vector-\Sigma}&=&\frac{7}{4}\, T(a)\,c^{(2)}_0\,,\nonumber\\
c^{(2)}_{\rm vector-scalar}&=&7\,T(r)\,c^{(2)}_0\,,
\end{eqnarray}
where we have included factors $\frac{1}{2}$ and $2$ to account for the 
reality of $\Sigma$ and for the presence of two hypermultiplet scalars
respectively. The two fermionic contributions are precisely as in the 
Eq.~(\ref{2loopgauge}), just with $T(r)$ replaced by $T(a)$ in the case of 
the gaugino. 

The contributions of the charged hypermultiplet thus sums up to
\begin{multline}
c^{(2)}_{\rm vector-scalar}
+c^{(2)}_{{\rm scalar}-\Sigma}
+c^{(2)}_{\rm scalar-scalar}
+c^{(2)}_{\rm vector-fermion}
+c^{(2)}_{\rm fermion-\Sigma}
+c^{(2)}_{\rm gaugino-fermion-scalar}\\
=\left( 7+2+3-3-1-8\right)T(r)\,c_0^{(2)}=0.\qquad\qquad\qquad\qquad{}
\end{multline}
Similarly, the self-interactions of the vector-multiplet give rise to
\be
c^{(2)}_{\rm vector}+
c^{(2)}_{\rm vector-\Sigma}+
c^{(2)}_{\rm vector-gaugino}+
c^{(2)}_{\rm gaugino-\Sigma}=
\left(\frac{9}{4}+\frac{7}{4}-3-1\right)T(a)\,c_0^{(2)}=
0\,.
\ee

Next, since we have kept the $d$ dependence in Eq.~(\ref{2loopgauge}), we
can immediately extend our analysis to a 4d ${\cal N}=1$ SYM theory. The 
2-loop Casimir energy contribution is proportional to 
\be
c^{(2)}_{\rm 4d~vector}+c^{(2)}_{\rm 4d~vector-gaugino}=(1-1)T(a)c_0^{(2)}
=0\,,
\ee
where the fermionic contribution includes an extra factor $\frac{1}{2}$ 
relative to Eq.~(\ref{2loopgauge}) to account for the chirality of the 
4d gaugino. 

Finally, consider the contribution of charged 4d matter.
The 2-loop effects based on gauge interactions can be inferred from
Eq.~(\ref{2loopgauge}) (with an appropriate factor $\frac{1}{2}$ for
fermion chirality in the loop).
The gaugino-fermion-scalar contribution is as in Eq.~(\ref{ss}), but
with an extra factor $\frac{1}{4}$ for the fermion chirality and
half the number of scalars. The scalar ``figure 8'' diagrams are
induced by the $D$ term potential and give rise to $\frac{1}{6}$ of
the contribution displayed in Eq.~(\ref{f8}) because of
the missing trace over $SU(2)_R$ degrees of freedom,
tr$\,\sigma_i\sigma_i=6$.\footnote{The scalar ``figure-8'' diagram
also contains a piece proportional to $(\operatorname{tr}{\bf T}^a)^2$
which is zero for nonabelian as well as anomaly-free abelian gauge
theories. The reason why the Casimir energy does not vanish in the
presence of an anomalous $U(1)$ is the associated occurrence of a
one-loop Fayet-Iliopoulos term which spontaneously breaks SUSY.}
Overall, one finds
\begin{multline}
c^{(2)}_{\rm 4d~vector-scalar}+c^{(2)}_{\rm 4d~scalar-scalar}+
c^{(2)}_{\rm 4d~vector-fermion}+c^{(2)}_{\rm 4d~gaugino-fermion-scalar}
\\
=\left(\frac{5}{2}+\frac{1}{2}-1-2\right)T(r)c_0^{(2)}=0\,,
\end{multline}
concluding our SUSY based checks.

\end{document}